\documentclass[mathleft,fleqn,%
]{an}

\usepackage{graphicx,fancyhdr}
\usepackage{rotating}

\usepackage{times}
\usepackage{natbib}

\begin{document}

\sloppy

\title{Data mining of near Earth asteroids in the Subaru Suprime-Cam archive\thanks{Based 
       on data collected at Subaru Telescope and obtained from the SMOKA, 
       which is operated by the Astronomy Data Center, National Astronomical 
       Observatory of Japan. The total data transfer disk space used for 
       this project was above 2.5 TB, being supported by Matei Conovici.}}

      \author{O. Vaduvescu\inst{1,2}\fnmsep\thanks{\email{ovidiuv@ing.iac.es}\newline}, 
              M. Conovici\inst{3},
              M. Popescu\inst{4,2,5},
              A. Sonka\inst{5,6},
              A. Paraschiv\inst{7,8},
              D. Lacatus\inst{7,8},
              A. Tudorica\inst{9,10},
              L. Hudin\inst{11},
              L. Curelaru\inst{12},
              V. Inceu\inst{13},
              D. Zavoianu\inst{5,12},
              R. Cornea\inst{12,14},
              R. Toma\inst{15,12},
              D.J. Asher\inst{15},
              J. Hadnett\inst{16,18},
              L. \'O Cheallaigh\inst{17,18}
              }

\titlerunning{Data mining of near Earth asteroids in the Subaru Suprime-Cam archive}
\authorrunning{O. Vaduvescu et al}

   \institute{Isaac Newton Group of Telescopes, 
              Apartado de Correos 321, E-38700 Santa Cruz de la Palma, Canary Islands, Spain 
            \and
              IMCCE, Observatoire de Paris, 
              77 Avenue Denfert-Rochereau, 75014 Paris Cedex, France 
            \and
              Amateur astronomer, Bucharest, Romania
            \and
              The Astronomical Institute of the Romanian Academy, 
              Cutitul de Argint 5, 040557 Bucharest, Romania
            \and
              Bucharest Astroclub, 
              B-dul Lascar Catargiu 21, sect 1, Bucharest, Romania 
            \and
              Admiral Vasile Urseanu Observatory, 
              B-dul Lascar Catargiu 21, sect 1, Bucharest, Romania 
            \and
              Institute of Geodynamics Sabba S. Stefanescu, 
              Jean-Louis Calderon 19-21, Bucharest, Romania, RO-020032, Romania
            \and
              Research Center for Atomic Physics and Astrophysics, Faculty of Physics, University of Bucharest, 
              Atomistilor 405, CP Mg-11, 077125 Magurele - Ilfov, Romania
            \and
              Bonn Cologne Graduate School of Physics and Astronomy, Germany 
            \and
              Rheinische-Friedrich-Wilhelms Universitaet Bonn, Argelander-Institut fur Astronomie, 
              Auf dem Hugel 71 D-53121 Bonn, Germany 
            \and
              Amateur astronomer, ROASTERR-1 Observatory, 400645 Cluj Napoca, Romania
            \and
              Romanian Society for Meteors and Astronomy (SARM), 130029 Targoviste, Romania
            \and
              Amateur astronomer, Cluj Napoca, Romania
            \and
              Babes-Bolyai University, 
              Faculty of Physics and Informatics, 400084 Cluj-Napoca, Romania 
            \and
              Armagh Observatory \& Planetarium, 
              College Hill, Armagh, BT61 9DG, Northern Ireland
            \and
              The Royal School, 
              College Hill, Armagh, BT61 9DH, Northern Ireland
            \and
              St Paul's High School, 
              108 Camlough Road, Bessbrook, Newry, BT35 7EE, Northern Ireland
            \and
              Visiting student, Armagh Observatory, 
              College Hill, Armagh, BT61 9DG, Northern Ireland
             }

\date{Submitted to Astronomische Nachrichten (Sep 2016); revised (Dec 2016)}

%\keywords{astrometry, minor planets, archives, data mining}

% -------------- Abstract

%\keywords{Minor Planets, Near Earth Asteroids, Data Mining, Image Archives, Orbital Amelioration, Distribution statistics}

\abstract {Abstract: As part of the EURONEAR project, almost 70,000 mosaic Suprime-Cam images taken 
           between 1999 and 2013 were data mined for about 9,800 near Earth asteroids (NEAs) known by 2013 May. 
           Using our PRECOVERY server and the new {\it Find Subaru CCD} tool, we scrutinized 4,186 candidate CCD 
           images possibly holding 518 NEAs. We found 113 NEAs as faint as $V<25$ magnitude, their positions being 
           measured in 589 images using Astrometrica, then reported to the Minor Planet Center. Among them, 18 
           objects represent encounters of previously single opposition NEAs, their orbital arcs being extended 
           by up to 10 years. 
           In the second part of this work we searched for unknown NEAs in 78 sequences (780 CCD fields) of 4-5 
           mosaic images selected from the same Suprime-Cam archive and totaling 16.6 deg$^2$, with the aim to 
           assess the faint NEA distribution observable with an 8-m class survey. A total of 2,018 moving objects 
           were measured, from which we identified 18 better NEA candidates. Using the $Rc$ filter in good weather 
           conditions, mostly dark time and sky directions slightly biased towards the ecliptic, at least one NEA 
           could be discovered in every 1 deg$^2$ surveyed. 
           } 

\maketitle

% -----------------------------------------------------------------

\section{Introduction}
\label{intro}

The continuous astrometric monitoring of near Earth asteroids (NEAs) and potentially 
hazardous asteroids (PHAs) is an important task for their orbital improvement and assessment of 
future risk of impact, as well as longer time follow-up necessary to study gravitational 
perturbations and other subtle effects such as the YORP and Yarkovsky \citep{vok15}, 
especially when very accurate astrometry is available. 

Within the EURONEAR project \citep{eur16a}, in 2007 we started to data mine some larger field 
image archives for known NEAs. As part of this project, \cite{vad09} implemented the first online 
application (known as ``PRECOVERY'') to search for the serendipitous encounters of all known 
NEAs and PHAs in one particular image archive, based on the IMCCE SkyBoT server \citep{ber06}. 
We applied this tool first to the Bucharest Observatory archive including 13,000 photographic 
plates. 

In our second similar project we data mined the Canada-France-Hawaii Telescope Legacy Survey 
(CFHTLS, numbering about 25,000 images), in which 143 known NEAs (including 27 PHAs) were found 
and measured in 508 images. As part of this project, 41 arcs were prolonged at their first or last 
opposition, 35 orbits were refined by adding new opposition data and 6 NEAs were recovered at their 
second opposition \citep{vad11a}. 

Our third project applied the same tool to data mine two larger field 2-meter class telescope 
archives located in the North (the ING/INT 2.5m) and South (the ESO/MPG 2.2m) comprising together 
330,000 images, finding 152 NEAs (including 44 PHAs) and reporting the resulting astrometry from 
761 images to the Minor Planet Center \citep{vad13a}. 

The present project and last in this suite applies PRECOVERY to an 8-meter class image archive, 
namely Subaru Suprime-Cam. This effort started at the end of 2011 and it was announced first in 
the ACM2012 meeting in Japan \citep{vad12}. 
In Section~\ref{camera-archive} we describe briefly the camera and its great survey capabilities, 
introducing also the SMOKA image archive. 
In Section~\ref{datamining} we recall the image reduction and search tools,
and their application to find and measure known NEAs.
Section~\ref{statistics} assesses the unknown NEA distribution at this faint level. 
Finally, Section~\ref{conclusions} draws the conclusions and proposes future plans. 

%__________________________________________________________________

\section{The Camera and Archive}
\label{camera-archive}

\subsection{Suprime-Cam on Subaru Telescope}

Installed in 1999 at the fast $F/1.86$ prime focus of the Subaru 8.2-meter national Japanese telescope 
located at 4,200~meters altitude atop Mauna Kea in Hawaii, the $34^\prime \times 27^\prime$ 80-mega pixel
Suprime-Cam CCD mosaic camera consists of ten CCDs of 4k$\times$2k ($4096 \times 2048$) pixels with a scale 
of $0.202^{\prime\prime}$ ($15\mu$ pixel size) in order to fit the excellent seeing at Mauna Kea - median 
value $0.61^{\prime\prime}$ in $i$ band \citep{miy02} matched by a study analyzing the first seven years 
actual Suprime-Cam PSF data \citep{nod10}. 

Thanks to the large aperture of the telescope (effective collecting area 51.65~m$^2$) and the large field 
of view of the prime focus camera, Subaru and Suprime-Cam offered the largest {\it etendue}\footnote{The 
{\it etendue} $A\Omega$ is defined as the product between the telescope light gathering power (effective 
aperture expressed in square meters) and the area of the sky imaged in a single exposure (deg$^2$).} 
in the world ($A\Omega=13.17$ m$^2\cdot$deg$^2$), being matched in 2006 by the larger field Pan-STARRS~1 
(similar etendue but $\sim3$ mag shallower), then in 2012 being surpassed by the DECam camera installed 
on the 4.2-meter Blanco telescope (etendue 25m$^2\cdot$deg$^2$ but $\sim1$ mag shallower than Suprime-Cam). 

In order to improve the quantum efficiency at redder wavelengths, in 2008 July Suprime-Cam was fitted 
with fully-depleted back-illuminated Hamamatsu Photonics KK (HPK) CCDs which replaced the old MIT/Lincoln 
Laboratory (MIT/LL) CCDs. The number of CCDs, their pixel size, plate scale and total field of view of the 
camera remained the same, the only change being the CCD numbering in the mosaic. We accommodate this change 
in our present work. 

\subsection{The SMOKA Image Archive}

Since 2002, SMOKA, acronym of the Subaru-Mitaka-Okayama-Kiso-Archive public science archive \citep{smo16} 
has provided access to the images and spectra observed with the Subaru national telescope plus other (mostly 
1-2~meter) telescopes of the Mitaka, Okayama, Kiso (University of Tokyo) and Higashi-Hiroshima observatories 
in Japan \citep{bab02}. 

A total of 81,878 Suprime-Cam raw science images have been incorporated by 2016 February 22 into the SMOKA 
archive. We used them to study some statistics, namely the distribution of sky pointings, and exposure times 
and filters used. 

\begin{figure}
\centering
\mbox{\includegraphics[width=8.5cm]{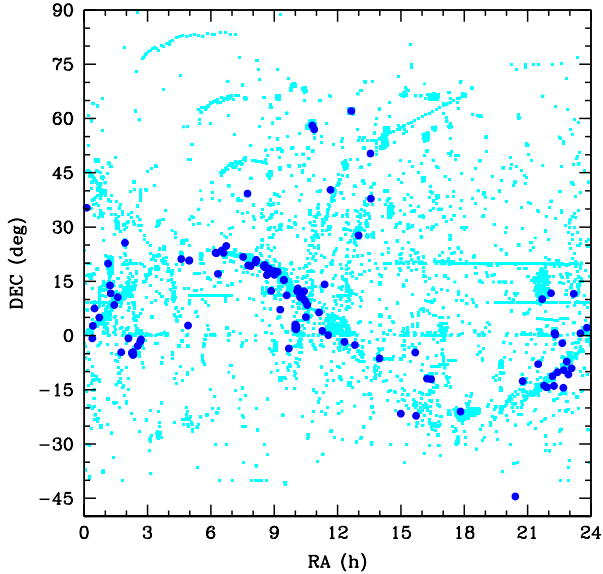}}
\begin{center}
\caption{The sky pointings of the Suprime-Cam archive (1999-2014: 81,878 images). The observed 
fields are plotted as small dots (cyan) and the NEAs (p)recovered in this project as larger dots (blue). 
}
\label{RADEC}
\end{center}
\end{figure}

Figure~\ref{RADEC} plots the sky pointings of the Suprime-Cam archive between 1999 January 5 and 2014 July 29 
(accessible by 2016 February 22). The observed fields are plotted as small dots (in cyan color). Most of the 
fields are distributed quite randomly on the sky, with the ecliptic covered by a few Solar System projects 
and other patterns representing mostly extragalactic projects. We overlay with dots (in blue color) the 
NEAs (p)recovered in this work (see Section \ref{foundNEAs}), located mostly close to the ecliptic. 

\begin{figure}
\centering
\mbox{\includegraphics[width=8.5cm]{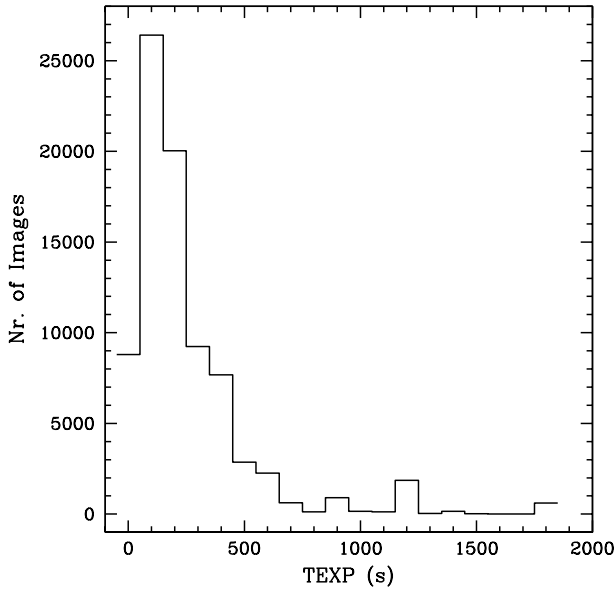}}
\begin{center}
\caption{The histogram of exposure times used by the Suprime-Cam archive (1999-2014: 81,878 images). 
The great majority of shots used exposures below 500s, and about half are shorter than 250s, allowing 
data mining of Solar system objects. 
}
\label{TEXP}
\end{center}
\end{figure}

Figure~\ref{TEXP} represents the distribution of the exposure times for the Suprime-Cam 1999-2014 archive 
(81,878 images). The great majority of the images used relatively short exposures (below 500~s, with about 
half below 250~s), which is feasible for data mining of NEAs and other Solar system objects, so that most 
trails remain small (a few pixels). 

Broad band filters were most popular ($86\%$) either in the Johnson-Cousins ($46\%$) or the Sloan system ($40\%$), 
while the intermediary band filters were used in only $2\%$ of cases, accounting together to $88\%$ of images 
feasible for data mining asteroids and other Solar system objects. The narrow band filters were used in 
$9\%$ of images, while other visitor filters (mostly narrow band) accounted for $3\%$. 

%__________________________________________________________________

\section{Data Mining the Subaru Suprime-Cam Archive}
\label{datamining}

In 2013 May almost 70,000 existing Suprime-Cam images (more exactly 69,333 observed between 1999 January 
and 2013 May) were searched for about 9,800 known NEAs (at that time) using the PRECOVERY server \citep{eur08}. 
We assumed for the search a safe $V=26$ limiting magnitude, possible to reach with Subaru in 100~s at $S/N=4$ detection 
level\footnote{Using the Subaru Imaging Exposure Time Calculator, http://www.naoj.org/cgi-bin/img\_etc.cgi} 
in dark conditions and good seeing, consistent with the proper motion and trailing loss effect for the large majority 
of NEAs. The search resulted in 4,186 candidate images possibly holding 518 NEAs. These findings include only 
asteroids encountered in at least two (typically 4-5) images of the same field taken at a short interval 
(typically less than 1~h), to allow image blinking needed to confirm the object's proper motion. 

\subsection{Image Reduction}

In a team of 10 people we used the SMOKA server to manually retrieve all the raw candidate mosaic images 
possibly holding the asteroids. Appropriate flat fields corresponding to the observing filters and dates were 
selected and downloaded from SMOKA, while the bias was taken from the overscan CCD regions of each science image. 
The raw science images (4,186 images of 10 CCDs each, totaling 700 GB) were reduced locally by Matei Conovici using 
the SDFRED Suprime-Cam software \citep{yag02,ouc04}, then posted on his private server (ca.\ 1,400 GB) for download
and carefully searched by our remotely distributed team. 

\subsection{Find Subaru CCD}

\begin{figure}
\centering
\mbox{\includegraphics[width=8.5cm]{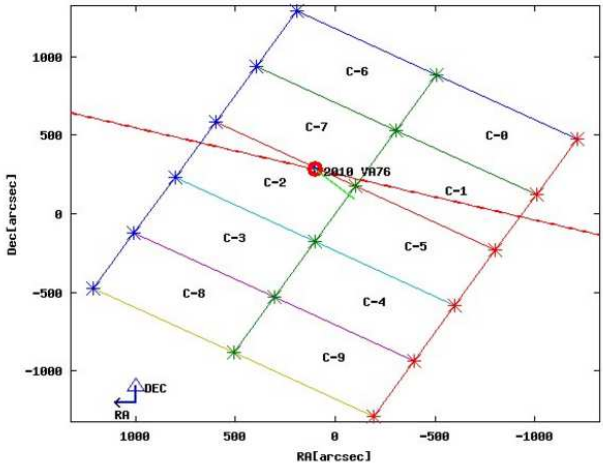}}
\begin{center}
\caption{Sample of the {\it Find Subaru CCD} output searching for NEAs in the field SUPA0057368 ($PA=120$ deg), 
overlaying the Suprime-Cam mosaic and the uncertainty (red crossing) line of the encountered object 2010~VA76. 
}
\label{FindSubaruCCD}
\end{center}
\end{figure}

To search for the CCDs possibly holding the asteroids, the dedicated tool {\it Find Subaru CCD} was written 
in PHP by Marcel Popescu and deployed on the EURONEAR website \citep{eur13}, which could be freely used for other 
asteroid Suprime-Cam data mining projects. Given the candidate image number (as reported by PRECOVERY) and the 
correct position angle\footnote{It was found that the camera position angle was not always recorded correctly in 
the Suprime-Cam image headers.} (upon checking for possible rotation), {\it Find Subaru CCD} plots all known NEAs 
in any observed Suprime-Cam field (using the SkyBoT server \citep{imc16}), overlaying the 10 CCD fields and the 
uncertainty region of poorly observed NEAs (obtained by querying NEODyS \citep{neo16}), so that the user can 
easily identify all CCDs possibly holding the object. Figure~\ref{FindSubaruCCD} plots one example run with 
the {\it Find Subaru CCD} output. 

\subsection{Found NEAs}
\label{foundNEAs}

We distributed the fields randomly, to be searched by a team of about 10 people (amateur astronomers and students,  
co-authors of this paper) who used the {\it Astrometrica} software \citep{raa16} to blink all candidate images. 
We searched the targets around their expected ephemerides (if the uncertainty $\sigma$ was small) or along the 
uncertainty regions predicted by NEODyS (if $\sigma$ was larger than $\sim10^{\prime\prime}$). 

A total of 113 known NEAs (plotted with large blue dots in Figure~\ref{RADEC}) were found in 589 images, 
representing only $22\%$ of the candidate objects, due to the high PRECOVERY threshold used $V=26$. Of these 
113 objects, 26 are PHAs, 95 corresponded to multiple opposition NEAs and 18 were one-opposition NEAs (poorly 
observed objects, including two PHAs). Most encounters resulted in small trails (typically under $\sim3\arcsec$)
whose centroids were easily measured by {\it Astrometrica}, even though most asteroids were slightly elongated 
into ellipses instead of circular sources. For longer trails we carefully visually measured the two ends which 
were averaged to report positions at standard mid-observed time. The astrometric reductions used the 
PPMXL reference system \citep{roe10}. The measured positions of the 113 known NEAs were reported to MPC 
between 2013 December and 2014 September. 

\begin{figure}
\centering
\mbox{\includegraphics[width=8.5cm]{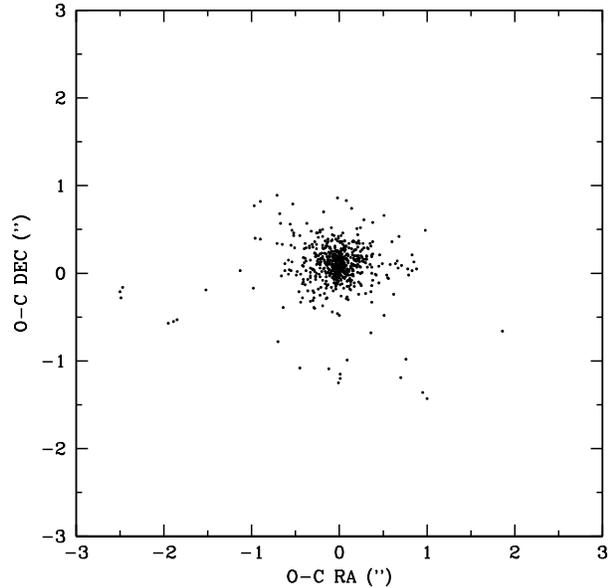}}
\begin{center}
\caption{The O$-$C (observed minus calculated) residuals for the 589 measured positions of 113 NEAs, 
calculated based on ephemerides derived from the improved orbits (2016 Feb 24). Most residuals are 
confined close to zero (standard deviations $0.37^{\prime\prime}$ in $\alpha$ and $0.27^{\prime\prime}$ 
in $\delta$), and only $2\%$ of points reside outside the $1^{\prime\prime} \times 1^{\prime\prime}$ box. 
}
\label{OminC}
\end{center}
\end{figure}

Figure~\ref{OminC} plots the O$-$C residuals (observed minus calculated) for all 589 measurements, 
obtained with our O$-$C calculator \citep{eur16b} querying very accurate NEODyS ephemerides based on the 
improved orbits (by 2016 February 24). Most of the points are confined around the origin, with standard 
deviation $0.37^{\prime\prime}$ in $\alpha$ and $0.27^{\prime\prime}$ in $\delta$. Only 16 points ($2\%$ 
of all data) sit outside $1^{\prime\prime}$ in either $\alpha$ or $\delta$, most of these measurements 
being affected by longer and fainter trails whose ends are more difficult to assess. 

Subsequently to our Suprime-Cam (p)recoveries and thanks to the greatly improved orbits, two objects 
were found in the SDSS archive by Lucian Hudin (2012~HC34 and 2010~DM21), being measured and reported 
to the Minor Planet Center as part of the same project. Also, following our data submission, some 
objects were data mined by other authors in other archives, being reported and then published together 
with our data in the same publication (e.g., 2012~HC34) or later. 

Table~\ref{table1} presents our 18 one-opposition recoveries. From these, the following six cases 
deserve special status, because these objects could have been lost without the Suprime-Cam recovery 
data. 2010~SZ3 was discovered by the Catalina survey in 2010 September, its one day arc being prolonged by 
us (Lucian Hudin) almost one month later at high uncertainty (about 2 degrees), after which it remains 
unobserved until today. 
2007~TK15 was discovered by Catalina in 2007 October and followed during one month, being precovered 
by Adrian Sonka 20 months before discovery, which allowed its recent recovery in 2015 at a very faint 
$V=23$ limit. 
2011~GM44 is a PHA discovered by the Catalina Siding Spring survey in 2011 April, observed during one month, 
and precovered by Adrian Sonka five years before discovery, then easily recovered recently in 2016. 
2011~KW19 was discovered by Pan-STARRS in 2011 May and observed for two months, precovered by Lucian 
Hudin seven years before discovery; it is unobserved since but the hugely reduced future sky plane
uncertainty allows easy recovery, for example in 2023.
2007~UA2 was discovered by Catalina in 2007 October, followed during four months, precovered by Lucian 
Hudin three years before, and is since unobserved but again is now easy to find in future (e.g., in 2022).
2002~VR14 is a very old NEA (not observed for 14 years) discovered by the NEAT survey in 2002 November,
followed during seven days only, and precovered by us (Ovidiu Vaduvescu) one month before discovery; the 
1-month arc will enable correct linkage when it is found again (e.g., by LSST).

\renewcommand{\arraystretch}{0.8}
\begin{table*}[!t]
\begin{center}	
\caption{One opposition NEAs recovered in the Subaru Suprime-Cam archive.
The sky plane uncertainty $\sigma$ at the time of (p)recovery is in arcsec.} 
\label{table1}
%%\resizebox{16.5cm}{!}{
\begin{tabular}{llrrrll}
\hline
\hline
\noalign{\smallskip}
\noalign{\smallskip}
Asteroid  &  Class  &  $\sigma$ ($^{\prime\prime}$) &  Nr. pos.  &   Arc (before/after) &   Reference  &    Reducers  \\
\noalign{\smallskip}\noalign{\smallskip}
\hline
\noalign{\smallskip}\noalign{\smallskip}
2012 HC34  &  NEA &  2200      &   10   &   6m/10y precovery  &   MPS 504077 &  L. Hudin \\ 
2010 SZ3   &  NEA &  6600      &   11   &    1d/1m recovery   &   MPS 504065 &  L. Hudin \\
2012 KC6   &  PHA &   140      &    6   &    2m/4y precovery  &   MPS 504077 &  D. Lacatus \\ 
2010 DM21  &  NEA &  2100      &   15   &    2m/7m precovery  &   MPS 505427, 504060, 505428 & M. Conovici, L. Curelaru \\
2009 UE2   &  NEA &    25      &    4   &    5m/2y precovery  &   MPS 505424 &  F. Ursache \\
2008 TJ157 &  NEA &    10      &   12   &  48d/52d precovery  &   MPS 505415 &  D. Lacatus, A. Paraschiv \\
2007 TK15  &  NEA &    76      &    3   &   1m/20m precovery  &   MPS 505407 &  A. Sonka \\
2011 GM44  &  PHA &  2247      &    4   &    1m/5y precovery  &   MPS 506465 &  A. Sonka \\
2008 UE202 &  NEA &    15      &    6   &  19d/30d precovery  &   MPS 505416 &  D. Lacatus, A. Paraschiv \\
2008 TZ    &  NEA &     1      &    6   &   8d/10d precovery  &   MPS 505415 &  D. Lacatus, A. Paraschiv \\
2011 KW19  &  NEA &   524      &    2   &    2m/7y precovery  &   MPS 506468 &  L. Hudin \\
2007 UA2   &  NEA &    38      &    3   &    4m/3y precovery  &   MPS 506380 &  L. Hudin \\
2008 BC22  &  NEA &    59      &    6   &    5m/3y recovery   &   MPEC 2014-Q72 &  D. Lacatus \\
2001 XP    &  NEA &    25      &    3   &   11d/1m precovery  &   MPS 528063 &  A. Tudorica \\
2006 QY5   &  NEA &   254      &    3   &    2m/5y precovery  &   MPEC 2006-Q15 &  A. Sonka \\
2002 VR14  &  NEA &    29      &    3   &    7d/1m precovery  &   MPS 528068 &  O. Vaduvescu \\
2001 HK31  &  NEA &     1      &    3   &  51d/59d precovery  &   MPS 528059 &  L. Hudin \\
2007 DD    &  NEA  &    2      &    7   &   14m/4y recovery   &   MPEC 2007-D15 &  L. Hudin \\
\noalign{\smallskip}
\hline
\hline
\end{tabular}
%%}
\end{center}
\end{table*}

\begin{figure}
\centering
\mbox{\includegraphics[width=8.5cm]{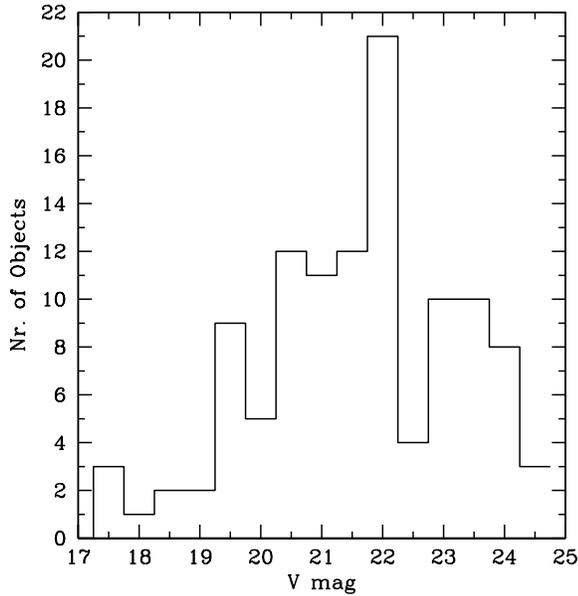}}
\begin{center}
\caption{Histogram of the V magnitudes (as given by Minor Planet Center ephemerides) 
of the 113 (p)recovered NEAs. 
Most objects were recovered between $V=19-24$ with a few faintest NEAs close to $V=25$. 
}
\label{MAG}
\end{center}
\end{figure}

Figure~\ref{MAG} presents the histogram showing the distribution of the $V$ magnitudes of the 113 
(p)recovered NEAs. The peak is around $V\sim22$, with the faintest objects recovered close to $V\sim25$, 
as expected based on the capabilities of Subaru. The faintest objects were 2007~UA2 ($V=24.6$ found by 
Lucian Hudin), 2007~TK15 and (283457) 2001~MQ3 (both at $V=24.8$ found by Adrian Sonka), with the first 
two among the above special cases. 

%__________________________________________________________________

\section{Statistics of the Faint NEA Distribution}
\label{statistics}

Using the entire Suprime-Cam archive existing by 2013, we assessed the NEA density 
observed with an 8-m class telescope in random directions and good weather conditions
(seeing around $0.8^{\prime\prime}$, according to \cite{nod10}). Using SMOKA we 
considered again all the Suprime-Cam images observed between 1999 January and 2013 May 
(69,333 mosaic images). 

\subsection{Sample Selection}

Based on the ASCII Suprime-Cam pointing archive alone, we selected all suitable 
``sequences'' defined as sets of 4 or 5 Suprime-Cam mosaic images having matching observation 
date and time (within 1 hour), telescope pointing (within maximum 2$^\prime$ dithering) 
and filter (accepting only the $Rc$-band images). No other conditions were imposed regarding 
the weather (seeing), Moon phase or distance, observed airmass, ecliptic latitude or 
Solar elongation. Using the first three criteria, we selected 108 sequences of 4-5 images, 
a total of 498 Suprime-Cam mosaic images for visual search and identification of moving 
sources. As the mosaic camera has 10 CCDs, there were potentially 1080 CCD sequences to search.

\subsection{Search for Moving Objects}

Matei Conovici used the SMOKA server to automatically retrieve all 1080 selected images. 
Appropriate flat fields corresponding to the $Rc$ filter and observing dates were selected and 
downloaded from SMOKA, while the bias was taken from the overscan CCD regions of each science 
image. The raw science images (498 images of 10 CCDs each, totaling 90 GB) were reduced locally 
by Matei Conovici using the same SDFRED Suprime-Cam software \citep{yag02,ouc04}, then posted 
on his private server (ca.\ 180 GB) and distributed for download by a team of 10 co-authors. 
We visually inspected all images, finally dropping 30 Suprime-Cam fields plus a few CCDs from 
6 other fields, owing to bad weather, bad seeing, shifted images, or nebulae producing a lack 
of enough astrometric stars. 

We carefully analyzed the remaining 78 Suprime-Cam fields which total 16.6 deg$^2$ 
on the sky (taking into account the dithers), measuring 2,018 moving objects (8,783 positions). 
To blink the images, identify all moving objects and obtain the astrometry, we used 
{\it Astrometrica} \citep{raa16} by loading all (4 or 5) images available for each CCD and matching 
the stationary sources with PPMXL catalog stars. Whenever {\it Astrometrica} did not work, we used 
first the {\it Astrometry.net} webtool \citep{lan09,lan10} to find the correct CCD centers and 
position angle (in some cases found inconsistent in the Suprime-Cam headers). 

\subsection{Search for NEA Candidates}

Having measured the astrometry of the 2,018 moving objects, we ran the Minor Planet Center's 
NEO Rating Tool \citep{mpc16}, identifying 141 objects with NEO scores more than $10\%$ (a very 
low threshold compared with the recommended $65\%$ value of the MPC). All these 141 higher scored 
objects were submitted to the MPC on 2016 September 5 (674 positions). To double check the actual 
number of NEA candidates, we considered these 141 candidates scored above $10\%$ against our own 
$\epsilon-\mu$ model \citep{vad11b} which is based on two observational quantities (the Solar 
elongation $\epsilon$ measured along the ecliptic and the proper motion $\mu$). 

\begin{figure}
\centering
\mbox{\includegraphics[width=8.5cm]{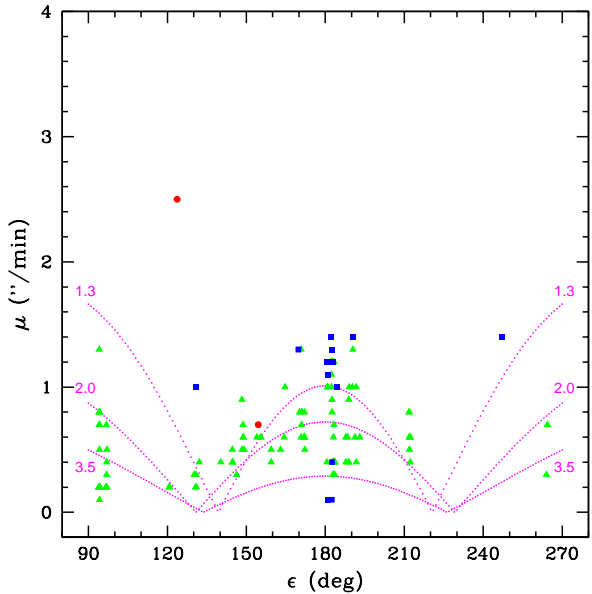}}
\begin{center}
\caption{The 141 higher scored objects (MPC NEO Rating scores above $10\%$) plotted against our 
$\epsilon-\mu$ model \citep{vad11b}. Using these two criteria, we believe there are at least 15 
real NEOs in the 78 Suprime-Cam analyzed fields (16.6 deg$^2$ total). Three very fast moving 
objects not shown (above top). 
}
\label{NEAcand}
\end{center}
\end{figure}

In Figure~\ref{NEAcand} we plot all the 141 higher scored objects, focusing on those located 
above the 1.3 a.u. NEA border (the upper curve plotted with magenta color). Three objects are
located above the plot (moving very fast between 5 and $25^{\prime\prime}$/sec) and left faint 
trails on images due to their fast motion: in these cases the trail ends were carefully measured 
and averaged to improve the accuracy. Red dots correspond to objects scored above $90\%$ by the 
MPC NEO Rating tool, blue squares to objects between $40\%$ and $90\%$, and green triangles to 
lowest scored objects between $10\%$ and $40\%$. 
Table~\ref{table2} includes the 18 NEA candidates with score above $40\%$ (red dots and blue squares 
in Figure~\ref{NEAcand}). We list our designation, observing date and time (mid of first image), 
the Suprime-Cam images (last digit representing the CCD number), the MPC NEO Rating, the 
ecliptic latitude $\beta$, Solar elongation $\epsilon$, proper motion $\mu$, magnitude and 
the field reducer. 

\renewcommand{\arraystretch}{0.8}
\begin{table*}[!t]
\begin{center}	
\caption{New NEA candidates with MPC NEO Rating above $40\%$ detected in the selected sequences
from the Subaru Suprime-Cam archive (78 fields). First group of 5 objects represents the best NEO candidates 
(MPC scores above $90\%$).
Moving object SDZV168 detected in two different sequences from same night.
Moon below horizon except for VUVb147 (34\% illuminated at distance 99\degr)
and SDZV189 (23\% at 69\degr).
}
\label{table2}
\begin{tabular}{@{}lllrrcrcl@{}}
\hline
\hline
\noalign{\smallskip}
\noalign{\smallskip}
             &                  &                  & \multicolumn{1}{l}{NEO} & & & & $Rc$ &  \\
Designation  &  Obs. date (UT)  &  Suprime-Cam image numbers & Rating ($\%$) & $\beta$ (\degr) & $\epsilon$ (\degr) & $\mu$ ($^{\prime\prime}$/min) & mag &   Reducer  \\
\noalign{\smallskip}\noalign{\smallskip}
\hline
\noalign{\smallskip}\noalign{\smallskip}
SAS0151  &  2001 10 21.58980  &  00067640 660 681         &   100   &    +2 & 183 & 25.3 & 23.0 & A. Sonka \\
SLH0213  &  2006 01 01.38699  &  00447847 877 907 937 967 &   100   &   +17 & 171 & 16.4 & 22.1 & L. Hudin  \\ 
VUVb147  &  2004 08 09.58900  &  00332646 676 700 730 760 &   100   &    +1 & 171 &  5.6 & 22.4 & V. Inceu \\
SDZV189  &  2004 01 17.61912  &  00267191 201 211 221 281 &    97   & $-$33 & 124 &  2.5 & 21.9 & D. Zavoianu \\
SATV071  &  2003 04 04.57860  &  00199203 213 233         &    97   &   +48 & 155 &  0.7 & 20.0 & M. Conovici \\
\noalign{\smallskip}\noalign{\smallskip}

SAS0364  &  2003 01 30.46192  &  00179729 739 749 759 769 &    84   &   +16 & 131 &  1.0 & 23.7 & A. Sonka \\
SLC0158  &  2001 01 25.35933  &  00034553 563 573 583     &    70   &    +6 & 183 &  1.2 & 19.6 & L. Curelaru \\
SLC0153  &  2001 01 25.35534  &  00034550 560 570 580 590 &    64   &    +6 & 183 &  0.4 & 20.1 & L. Curelaru \\
VDA1004  &  2011 05 05.34111  &  01314465 475 485 495 505 &    59   &   +35 & 247 &  1.4 & 20.7 & D. Zavoianu \\
SDA1001  &  2011 05 05.57912  &  01314946 956 966 976 986 &    57   &   +30 & 184 &  1.0 & 23.1 & D. Zavoianu \\
SLH0110  &  2002 09 03.33220  &  00120162 192 222 252 282 &    56   &  $-$1 & 182 &  1.4 & 23.8 & L. Hudin \\
SUVI028  &  2002 09 03.54191  &  00121033 063 093 123 153 &    51   &  $-$1 & 183 &  1.3 & 23.6 & V. Inceu \\
SDZV096  &  2002 09 02.33937  &  00118824 854 884 914 944 &    49   &    +1 & 181 &  1.2 & 22.5 & D. Zavoianu \\
SLO0037  &  2010 06 11.27865  &  01228086 096 106 116     &    48   &  $-$3 & 191 &  1.4 & 22.3 & L. \smash{\'O} Cheallaigh \\
SLH0122  &  2002 09 03.33220  &  00120163 193 223 253 283 &    45   &  $-$1 & 183 &  0.1 & 22.9 & L. Hudin \\
SAS0063  &  2002 09 02.42399  &  00119204 234 264 294 324 &    42   &    +1 & 181 &  1.1 & 23.9 & A. Sonka \\
SDZV168  &  2002 09 02.27683  &  00118559 589 619 649 679 &    41   &    +1 & 181 &  0.1 & 22.2 & A. Sonka \\
...      &  ...               &  00118829 859 889 919 949 &   ...   &   ... & ... &  ... & ...  & D. Zavoianu \\
SLO0011  &  2005 12 31.39525  &  00445826 836 846 856 866 &    40   &   +17 & 170 &  1.3 & 21.7 & L. \smash{\'O} Cheallaigh \\
\noalign{\smallskip}
\hline
\hline
\end{tabular}
\end{center}
\end{table*}

%______________________________________________________________

\subsection{NEA Sky Density and Comparison with Past Work}

In a similar study, applying both the $\epsilon-\mu$ model and MPC NEO Ratings to 47 
known NEAs, \cite{vad13b} found that at least three quarters are quite clearly identified 
as NEAs, with several more being marginally identifiable as NEAs using these criteria, and 
a few being impossible to separate from main belt asteroids (MBAs) based on a single night's 
observation. The converse question is of false positives, whether MBAs can appear slightly 
above the 1.3 a.u. border (Figure~\ref{NEAcand}) which we use to highlight NEAs. While the 
objects a long way above that border are unambiguously NEAs, our statistic of total NEAs 
found in these Suprime-Cam sequences 
depends quite strongly on whether most of our several suspected NEAs visible around 
$\epsilon \sim 180\degr$ in Figure~\ref{NEAcand} are indeed NEAs. Sky motion is an 
especially good NEA discriminator near opposition \citep{jed03}, and our own random trial 
to check sky motions of known asteroids, normalizing to $\sim$1,300 with $170 < \epsilon < 190\degr$ 
(1,200+ of our 2,018 unknown Suprime-Cam objects being within this range), yielded only one 
non-NEA around $\mu=1.4\arcsec$/min and three more just above the 1.3 a.u. border but below 
1.2\arcsec/min. 

Based on the above and on our experience from other projects regarding follow-up of similar 
NEA candidates observed with the Isaac Newton Telescope (INT) and other telescopes from the 
EURONEAR network \citep{vad11b,vad13b,vad15}, in Figure~\ref{NEAcand} we estimate at least 15 
new NEAs identified by our team in the analyzed 78 Suprime-Cam fields. 

\begin{figure}
\centering
\mbox{\includegraphics[width=8.5cm]{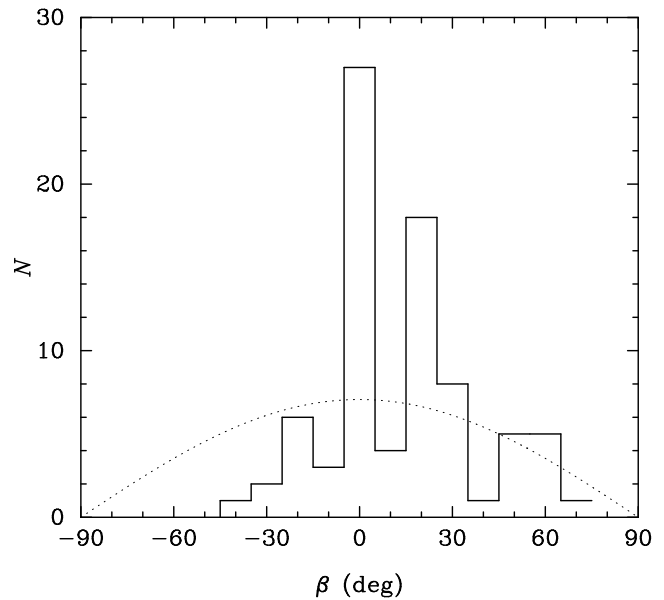}}
\begin{center}
\caption{Ecliptic latitudes of the 78 image sequences that were suitable for searching. There 
  is some concentration towards low latitudes compared to an isotropically random distribution 
  (dotted line) 
  but not the exclusive choice of near-ecliptic fields that characterizes some Solar system surveys. 
}
\label{beta}
\end{center}
\end{figure}

This result of at least 15 NEAs encountered in the total covered field of 16.6 deg$^2$
allows us to conclude that using the Suprime-Cam with the $Rc$-band filter, 
at least one NEA could 
be found in 1.1 deg$^2$ or $\sim$4 Suprime-Cam fields observed in random directions 
(ecliptic latitude distribution in Figure~\ref{beta}). Most image sequences ($85\%$) were 
obtained in dark time, and only $15\%$ with gray Moon typically at low altitude. 

Among the 18 best NEA candidates (Table~\ref{table2}), none are beyond $Rc$=24.0 
but several are beyond 23.0 and several more beyond 22.0 magnitudes. The faint object 
detectability is similar to Figure~\ref{MAG} but many brighter NEAs are already known 
rather than waiting to be found as unknown.  \cite{ter13} obtained near-completeness 
just beyond $r=24.0$ for high latitude MBAs with Suprime-Cam: trailing loss would 
remove a few NEAs that have higher sky motions. 

In a similar teamwork survey covering 24 deg$^2$ with 2-m class telescopes capable to
reach limiting magnitude $V\sim23$ (ESO/MPG and ING/INT), \cite{vad11b} found that on average 
one NEA could be observable scanning randomly 2 deg$^2$ of dark sky. Later, using only INT 
2.5m data covering 44 deg$^2$, \cite{vad15} concluded that in dark conditions one NEA could 
be discovered in at least 2.8 deg$^2$. 

None of our best 7 NEA candidates found with the 4.2m Blanco telescope capable to reach 
$V\sim24$ was fainter than $R=22.5$ though we found a small number of likely MBAs fainter than 
23.0 mag \citep{vad13b}. Our conclusion from those limited data was one NEA per $\sim$1.4 deg$^2$, 
possible to discover with a 4-m class telescope. 

In 2001 February and October, \cite{yos03} and \cite{yos07} used the Suprime-Cam for two 
small surveys around opposition (covering 3 and 4 deg$^2$, respectively) to investigate the 
very small MBA populations (sub 1-km, up to limiting apparent magnitudes $R=25-26$). 
Based on their past work, their actual hypothesis is that about one NEA could be found in every 
Suprime-Cam field (Fumi Yoshida - private communication), which apparently exceeds our present findings 
by 4 times. However, their surveys were conducted at opposition (within $\pm3$\degr\ from the 
ecliptic) and in dark conditions, in comparison with our randomly selected fields covering 
various ecliptic latitudes. 

The ecliptic latitude distribution of our sample (Figure~\ref{beta}) probably affected 
our findings, as most NEAs are found closer to the ecliptic \citep{ray04,ver09}. Our Blanco 
likely NEA candidates \citep{vad13b} showed the same pattern, with 6 of the 7 objects having 
$\beta$ within $\pm$5\degr\ despite only about half the fields searched being in that range. 
The $\beta$ values in Table~\ref{table2} suggest that near the ecliptic at least one NEA 
per 0.5 deg$^2$ could be discovered with Subaru/Suprime-Cam. 

\subsection{Search of Possible Pairings with Virtual Impactors}

We tested the entire known Virtual Impactor (VI) population, namely 631 bodies available from the 
NASA/JPL Sentry Risk Table \citep{nas16} for possible pairings with our 141 Subaru NEA candidates. 
The automatic PHP tool $VICheck$ was built by Marcel Popescu, to test all possible combinations 
(631 VIs $\times$ 141 candidates = 88,971 combinations) using the MPC Orbits/Observations database 
and our Subaru observations, running the $FO$ batch orbital fit software provided with the $Find\_Orb$ 
package by \cite{gra16}. After almost 12 days running on a typical Linux desktop Intel(R) Core(TM) 
i7-2600K CPU 3.40GHz, followed by manual check using $Find\_Orb$ of about 100 possible pairs (defined 
as generating small orbital RMS - few arcsec after fitting a given VI and Subaru pair using $FO$), 
no link between any known VI and our Subaru NEA candidates could be found.

\section{Conclusions and Future Work}
\label{conclusions}

This represents the fourth data mining project carried out within the EURONEAR project with the contribution 
of students and amateur astronomers. We used the Subaru SMOKA archive with two aims, searching the database of 
almost 70,000 Suprime-Cam mosaic images taken between 1999 January and 2013 May. 

First, we searched for all known NEAs serendipitously falling in the Suprime-Cam archive images, 
in order to improve their orbits, especially those poorly observed. 
Our EURONEAR PRECOVERY server identified 4,186 candidate images potentially holding 518 NEAs, carefully 
checked by our team using the new tool {\it Find Subaru CCD} which overlays the NEAs and their uncertainties 
over the Suprime-Cam CCD mosaic layout to identify exactly the regions to be searched. This search yielded 
113 NEAs found, as faint as $V<25$ magnitude, which were measured with {\it Astrometrica} in 589 images 
and reported to the Minor Planet Center. Among these findings, 18 cases represent observations of 
previously single opposition NEAs, orbital arcs being extended by up to 10 years. 

Second, we searched for unknown moving objects in 78 sequences (780 CCD fields) of 4-5 mosaic images selected from 
the entire Suprime-Cam archive and totaling 16.6 deg$^2$, in order to assess the faint NEA distribution 
accessible to an 8-m class survey. From the total number of 2,018 measured moving objects, the use of 
two rating tools identified 18 better NEA candidates and a further 123 lower scored objects. Using the $Rc$ 
filter in good weather conditions, mostly dark time and sky directions slightly biased towards the ecliptic, 
we conclude that at least one NEA could be discovered in every 1 deg$^2$ surveyed (equivalent to 4 Suprime-Cam 
fields). This is an average of one NEA every 0.5 deg$^2$ near the ecliptic and a lower NEA density elsewhere. 

As part of the EURONEAR project, a moving object processing software (MOPS) pipeline is being developed 
by a Romanian PhD student, soon being tested and compared with human detection using archival images and a 
planned asteroid mini-survey using the 2.5m Isaac Newton Telescope (INT) in La Palma. Looking further, the 
new Hyper Suprime-Cam (HSC) mounted in 2013 at the same prime focus of the Subaru telescope currently represents 
the world's largest survey facility, covering an {\it etendue} $A\Omega=91.4$ m$^2\cdot$deg$^2$ which surpasses 
by 7 times that of Suprime-Cam and Pan-STARRS 1, and by 3 times the Blanco-DECam {\it etendue}. 
Our {\it Mega-Precovery} server \citep{eur12} now accesses the entire Suprime-Cam archive (part of the overall 
SMOKA archive collection), and has started to include the HSC images in its {\it Mega-Archive}. They could be
searched for any known NEAs or other Solar System objects. Before the LSST era, we hope to use the HSC for an 
NEA mini-survey, covering selected ecliptic latitudes and solar elongations. This will allow us to test and 
expand our statistics about the faint NEA distribution observable with an 8-m class survey.

\smallskip
\begin{acknowledgements}

The paper is based on data collected at Subaru Telescope and obtained from the SMOKA, 
which is operated by the Astronomy Data Center, National Astronomical Observatory of Japan. 
The first author thanks Dr. Fumi Yoshida and Prof. Tsuko Nakamura for their encouragement 
received in the first stage of the project. 
Part of the work of M. Popescu was supported by a grant of the Romanian National Authority 
for Scientific Research – UEFISCDI, project number PN-II-RU-TE-2014-4-2199. 
Liam \'O Cheallaigh is grateful to Sentinus for the Nuffield Research Placement which enabled his work 
at Armagh Observatory. Research at Armagh is supported by the N. Ireland Dept.\ for Communities.
Other contributors were: Saoirse Doyle and Ryan Connelly (Armagh student visitors), Dan Vidican, Costel 
Opriseanu, Mihai Dascalu (Bucharest Astroclub), Toma Badescu (Romanian student in Bonn), and more 
recently Felician Ursache, Stoian Andrei and Andrei Marian Stoian (Romanian amateur astronomers). 
Many thanks to Dr. Aswin Sekhar who served as referee, suggesting some points which allowed us 
to improve our paper. 

\end{acknowledgements}

\end{document}